\documentclass[prb,twocolumn,showpacs,preprintnumbers,superscriptaddress,amsmath,amssymb]{revtex4}

\usepackage{epsf,color,graphicx}

\usepackage[dvips]{epsfig}

\begin{document}


\title{Photon lasing in a GaAs microcavity: similarities with a polariton condensate}



\author{ Daniele Bajoni, Pascale Senellart, Aristide Lema\^{i}tre and Jacqueline Bloch}
\affiliation{CNRS-Laboratoire de Photonique et Nanostructures, Route
de Nozay, 91460 Marcoussis, France}

\email[]{jacqueline.bloch@lpn.cnrs.fr}

\date{\today}

\begin{abstract}
We study experimentally the lasing regime of a GaAs based
microcavity sample under strong optical pumping. The very same
sample exhibits the strong coupling regime at low excitation power
with a Rabi splitting as large as 15 meV. We show that some features
which may be considered as experimental evidence of polariton Bose
Einstein condensation are also observed in the weak coupling regime
when the cavity is behaving as a regular photon laser. In
particular, the emission pattern in the lasing regime displays a
sharp peak near the energy minimum followed by a Boltzmann
distribution at higher energies.

\end{abstract}

\pacs{71.36.+c, 78.55.Cr, 78.45.+h}

\maketitle

Microcavity polaritons are the quasi-particles arising from the
strong coupling regime between a cavity mode and quantum well
excitons\cite{Weisbuch92}. Because of their bosonic nature,
microcavity polaritons are expected to undergo a Bose-Einstein
condensation with the appearance of spontaneous coherence
\cite{Livrekavokin,Keeling2007}. The polariton dispersion presents a
pronounced energy trap close to the center of the Brillouin zone
\cite{Houdre94}. As a result, polaritons exhibit a very small
effective mass ($\sim 10^{8}$ smaller than the hydrogen atom mass)
and thus are expected to condensate at unusually high temperatures
(up to room temperature in wide band gap
microcavities\cite{Malpuech2002}).

Claims of polariton condensation have been published these last
years by several groups in different cavity geometries and in
different semiconductor systems. In planar cavities, strong evidence
for the spontaneous appearance of polariton spatial coherence has
been obtained in II-VI CdTe based samples \cite{Kasprzak2006} at low
temperature (typically 10 K). More recently, polariton lasing has
been claimed at room temperature in wide band-gap GaN based
microcavities \cite{Christopoulos2007} but definite proof of the
persistence of the strong coupling regime in these measurements
still needs to be established. Concerning planar GaAs based
microcavities, it is now well established that polariton lasing
cannot be obtained under high energy non resonant
excitation\cite{Kira97,Pau97,Bloch2002,Butte2002}: because of a
relaxation bottleneck \cite{Tartakovskii2000,Senellart2000},
polaritons do not scatter efficiently enough toward the lowest
energy states and eventually the strong coupling regime is bleached
before polariton state occupancy in the energy minimum reaches
unity. To circumvent this inefficient polariton thermalization,
"cold" excitation has been successfully used by directly injecting
polaritons on the high energy states of the lower polariton branch
\cite{Deng2002}. Finally very recently Babili et al.
\cite{Snoke2007} reported on a strong non-linear coherent emission
in a GaAs based microcavity under high energy non resonant
excitation using a spatially localized energy trap induced by
strain.

Because at high excitation power, conventional lasing (VCSEL) can
occur in a semiconductor microcavity, it is delicate to ensure that
the observed emission is actually due to polaritons, especially in
new materials or when introducing a new geometry.

In the present work, we want to underline that striking similarities
exist between a polariton laser and a standard VCSEL. We perform
emission measurements under high energy non resonant excitation in a
large Rabi splitting sample analogous to the one used in Refs.
\onlinecite{Deng2002} and \onlinecite{Snoke2007}. Strong non-linear
emission is observed which is shown to be due to electron-hole pair
lasing by monitoring the in-plane dispersion of the emission.
Nevertheless we show that the lasing emission does not always occur
at the expected energy of the bare cavity mode but it is
significantly redshifted. Moreover, monitoring the emission as a
function of energy, we find an intensity distribution completely
similar to that reported for a polariton laser. Therefore, these
experimental features cannot be used on their own as a proof for the
establishment of a polariton condensate in coexistence with a
thermalized polariton cloud.

Our sample, grown by molecular beam epitaxy, consists in a
$\lambda/2$ AlAs cavity surrounded by two $Al_{0.2}Ga_{0.8}As/AlAs$
Bragg mirrors with respectively 16 and 20 pairs on the top and
bottom mirror. To get a large Rabi splitting with a small cavity
volume, we inserted quantum wells not only in the cavity layer but
also at the first antinodes of the electromagnetic field in each
Bragg mirror as firstly reported in Ref. \onlinecite{Bloch1998}. The
present sample contains three sets of 4 quantum wells, one at the
center of the cavity and one in each Bragg mirror. The epitaxial
layers (both cavity and mirrors) present a thickness gradient along
the wafer so that the relative energy between the cavity mode and
the exciton can be changed by moving the laser spot on the sample.
We define the detuning as $\delta = E_{C}(k=0) - E_{x}(k=0) $,
$E_{C}(k=0)$ and $E_{x}(k=0)$ being respectively the energy of the
cavity mode and of the heavy hole (HH) exciton at in-plane wavector
$k=0$. Photoluminescence (PL) experiments are performed using a cw
Ti:Saphire laser focused onto a 50 $\mu m$ diameter spot on the
sample with a 30 mm focal lens and a 50$^{\circ}$ angle of
incidence. The emission is collected through a 50 mm focal lens,
angularly selected with a 0.5$^{\circ}$ angular resolution,
spectrally dispersed with a double monochromator and detected with a
Si avalanche photodiode. For all measurements, the cavity sample is
held at 4K in a cold finger cryostat.

\begin{figure}[]
\includegraphics[width=0.9\columnwidth]{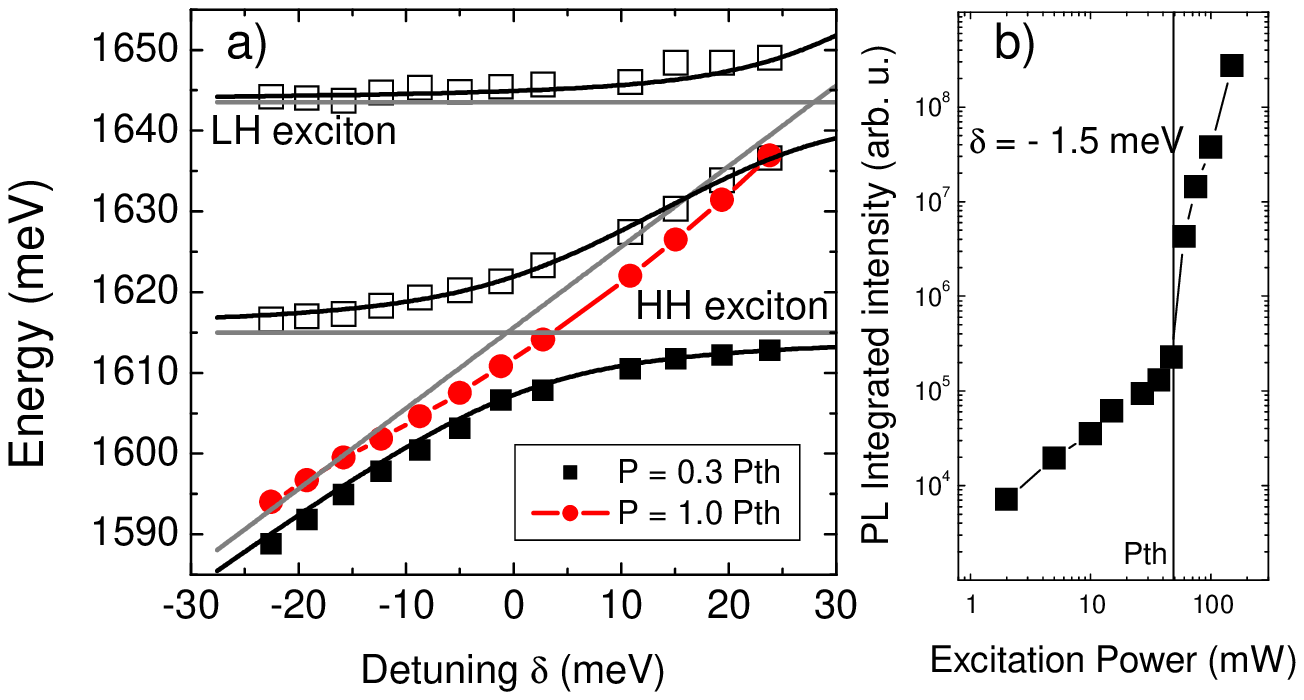}
\caption{(Color online) \textbf{(a)} squares : lower polariton
branch energy measured by photoluminescence around normal incidence
as a function of the cavity-exciton detuning $\delta$, open squares:
middle and upper polariton energy measured by photoluminescence
excitation spectroscopy of the lower polariton as a function of
$\delta$, thick black line: fit of the three polariton branches,
thin grey lines : deduced energy of the HH and LH exciton as well as
of the uncoupled cavity mode; red circles : energy of the emission
at threshold as a function of $\delta$; \textbf{(b)}Integrated
intensity measured around normal incidence as a function of the
non-resonant excitation power for $\delta = - 1.5 meV$ }\label{Fig1}
\end{figure}

Fig.1\textbf{(a)} presents the energy of the lower polariton branch
measured as a function of the detuning $\delta$ in the low density
regime. The same graph also presents the energy of the middle and
upper polaritons measured by photoluminescence excitation
spectroscopy of the lower polariton branch. Two anticrossings are
clearly observed between the cavity mode and the HH and light-hole
(LH) exciton. The energy of the three polariton branches is well
reproduced using a Rabi splitting of $\Omega_{HH}= 15\, meV$ and
$\Omega_{LH} = 12.5\, meV$ with the HH and LH exciton respectively.
Notice that the energy of the uncoupled cavity mode is precisely
determined in the anti-crossing region by monitoring the spectral
shift of the first reflectivity minimum on the high energy side of
the mirror stop-band.

\begin{figure}[]
\includegraphics[width=7 cm]{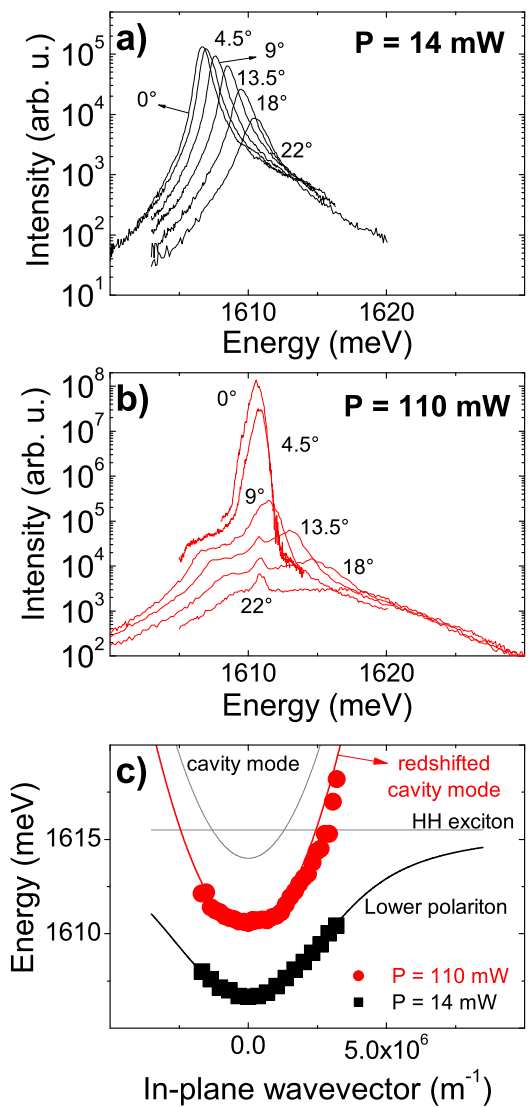}
 \caption{ (Color online) Photoluminescence spectra measured for different detection angles for an excitation power \textbf{(a)}
$P = 14\, mW$, \textbf{(b)} $P = 110\, mW$; \textbf{(c)} Emission
energy as a function of the in-plane wavector (deduced from the
detection angle) (squares) for $P = 14\, mW$ (circles), and $P =
110\, mW$. The black line corresponds to the calculated lower
polariton branch, the thin grey lines to the deduced uncoupled
cavity mode and HH exciton, and the thick red line to the cavity
mode redshifted by $3\, meV$. For the three figures, $ \delta = -
1.5 \, meV$ }\label{Fig2}
\end{figure}

Photoluminescence measurements as a function of the excitation power
were performed for various detunings under non-resonant excitation.
For each detuning the laser energy is tuned to the first
reflectivity minimum above the bragg-mirror stop-band, typically 120
meV above the polariton emission energy. A typical curve showing the
power dependence of the integrated intensity measured at k=0 is
shown in Fig. 1\textbf{(b)} for $\delta = - 1.5$ meV. Above a
threshold power $P_{Th}$, a very strong non-linear increase of the
emission intensity is observed. To check whether the system stays in
the strong coupling regime near threshold, we have performed angle
resolved photoluminescence for different excitation powers. Fig.
2\textbf{(a)} presents photoluminescence spectra measured at low
excitation power for several detection angles. The emission energy
as a function of the deduced in-plane wave-vector \cite{Houdre94} is
summarized in Fig. 2\textbf{(c)}. At low excitation power (0.3
$P_{th}$), the typical lower polariton dispersion is observed. Fig.
2\textbf{(b)} shows PL spectra measured at the same point of the
sample but for a higher excitation power (2.4 $P_{Th}$). At zero
degrees, the emission spectrum presents a single intense line
centered around 1611 meV. This line at 1611 meV is still visible at
larger angle but with a reduced intensity: this is due to Rayleigh
scattering within the sample. The emission spectra at finite
detection angles present an additional emission line at higher
energy than 1611 meV, continuously blueshifting with the angle of
detection. The measured dispersion of this line is summarized in
Fig. 2\textbf{(c)}. It is much steeper than the lower polariton
branch at low excitation density. It can be well fitted using the
cavity mode dispersion but rigidly red-shifted by roughly 3 meV.
These angle resolved photoluminescence measurements show that the
emission presents the energy dispersion of a pure photonic mode thus
indicating that the strong coupling regime is lost for this range of
excitation power.

Notice that the spectra of Fig. 2\textbf{(b)} also present a
shoulder at energies smaller than 1611 meV, exhibiting a small
blueshift with the angle of detection. This shoulder is due to
polariton emission coming from the edge of the detection spot, where
the carrier density is smaller and the strong coupling regime still
exists. Indeed, given our angular resolution $\Delta\vartheta =
0.5^{\circ}$, the diameter $d$ of the detection spot on the sample
surface has a lower limit $d_{min}$ given by diffraction:
\begin{displaymath}
d_{min}=0.61\frac{\lambda}{\tan\left(0.5\cdot\Delta\vartheta\right)}
\end{displaymath}
where $\lambda$ is the emission wavelength in vacuum. This gives
$d_{min} = 110$ $\mu$m, which is two to three times the dimension of
the excitation spot. This explains why polaritons in strong coupling
regime coming from the edge of the excitation spot are also visible
in the spectra of Fig. 2\textbf{(b)}.

\begin{figure}[]
\includegraphics[width=0.7\columnwidth]{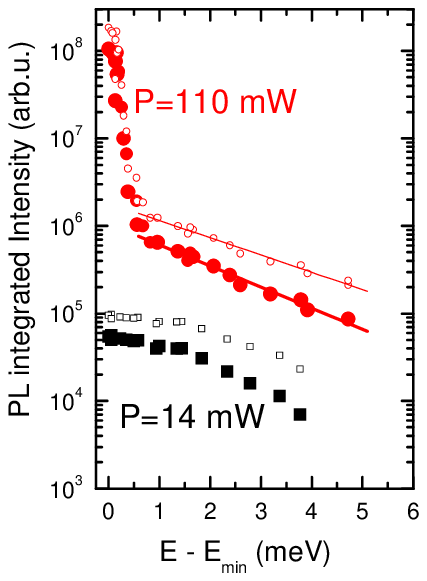}
\caption{(Color online) Full symbols: Photoluminescence integrated
intensity as a function of the emission energy for (black squares)
$P = 14 mW$, (red circles) $P = 110 mW$; Open symbols:
Photoluminescence integrated intensity divided by the square of the
polariton photonic Hopfield coefficient as a function of the
emission energy for (open black squares) $P = 14 mW$, (open red
circles) $P = 110 mW$; For each excitation power, $E_{min}$
corresponds to the emission energy measured at $k=0$. $ \delta = -
1.5 \, meV$}\label{Fig3}
\end{figure}

The energy of the emission line at threshold is plotted as a
function of detuning on Fig. 1\textbf{(a)}. For strong positive or
negative detunings, the emission energy at threshold matches that of
the bare cavity mode deduced from fitting the polariton
anticrossings. Nevertheless close to zero detuning or for positive
detunings up to $+10\, meV$, the emission line at threshold lies at
significantly lower energy than the calculated uncoupled cavity mode
deduced from low density measurements. This reduced blueshift of the
emission line when increasing excitation power could be mistaken for
a sign that the system remains in the strong coupling regime, with a
reduction of the Rabi splitting.

This incomplete blueshift, when lasing emission occurs, is clearly a
feature occurring close to the excitonic resonance. This can be
understood by considering the change of index of refraction of the
GaAs when evolving from the excitonic regime toward the
electron-hole plasma regime. A calculation of the low temperature
GaAs index of refraction has been reported in ref.
\onlinecite{Lowenau82} for several carrier densities. At low carrier
density, the real part of the index of refraction presents a
derivative-shaped excitonic feature centered at the exciton
resonance energy $E_{x}$, superimposed to a slowly varying function
$\widetilde{n}(E)$ slightly increasing with the energy.
$\widetilde{n}(E)$  accounts for the index of refraction due to all
other resonances in the quantum well, and is the index of refraction
value to be taken into account to calculate the energy of the
uncoupled cavity mode. For high carrier densities, the excitonic
feature progressively vanishes and the index of refraction
$n_{HD}(E)$  presents a peak at an energy $E_{HD}$ slightly higher
than the exciton energy, superimposed to a slowly varying function
close to $\widetilde{n}(E)$. Thus for energies close to $E_{HD}$,
$n_{HD}(E)$ is larger than $\widetilde{n}(E)$. As a result, the
cavity mode wavelength is at lower energy than what is deduced from
low density measurements. Thus the overall emission blueshift with
increasing excitation power is reduced. This incomplete blueshift
occurs only near $E_{HD}$, i.e. for zero or positive detunings (see
fig. 1a). It is induced by the refractive index of the QW layer only
and is not influenced by the rest of the cavity layers. It is thus
proportional to the overlap $\mathcal{A}$ between the
electromagnetic field of the cavity mode and the QW layer. Since the
Rabi splitting is proportional to $\sqrt{\mathcal{A}}$, the spectral
distance between the lasing emission and the calculated cavity mode
varies as $\Omega^{2}$. It is thus particularly pronounced in our
sample because it exhibits a very large Rabi splitting, as compared
to samples of previous reports\cite{Butte2002,Kira97} where the
transition toward the weak coupling regime was studied.

Finally let us describe the emission pattern of the present
microcavity. Fig. 3 summarizes the integrated intensity of the angle
resolved measurements presented in Fig. 2. The integrated intensity
is plotted as a function of the emission peak energy. In Fig. 3 we
also plot with open symbols the integrated intensity divided by the
square of the polariton Hopfield coefficient corresponding to the
polariton photon content. In the strong coupling regime, this
quantity is directly proportional to the polariton population. At
low excitation power, we find as in many previous works
\cite{Butte2002,Kasprzak2006} that the polariton population is not
thermalized. This is because the polariton-polariton interaction
time is much longer than the polariton lifetime. Above the lasing
threshold in the weak coupling regime, the intensity distribution
presents a pronounced peak close to k=0 (corresponding to $E =
E_{min}$) and an exponential decay at higher energy. This shape
remains very similar when dividing the intensity by the Hopfield
coefficient (red open symbols) even if this operation is meaningless
in the weak coupling regime. This emission pattern presents striking
similarities with that reported for the polariton condensate
\cite{Kasprzak2006,Snoke2007}. Such emission pattern has been
highlighted as a proof that a thermalized condensate is achieved.
This is an analogy to the case of atomic physics where a large
fraction of the atoms is in the condensate, and coexists with a
thermalized cloud of uncondensed atoms\cite{Ensher96}. In our case,
the system is in the photon lasing regime and the emission pattern
probably simply reflects the energy distribution of the
electron-hole pairs, the emission of which is filtered by the cavity
mode.

Our aim in this paper is to underline that this emission pattern, a
sharp peak near the energy minimum followed by a Boltzmann
distribution at higher energies, can be similarly observed in a
conventional laser in weak coupling regime. Although massive
occupation of the lower energy states is an important characteristic
of condensates, it cannot be used by itself to distinguish polariton
condensation from photon lasing.

To conclude we have studied a III-V GaAs based semiconductor
microcavity presenting a very small cavity volume and a large number
of quantum wells, thus being an ideal candidate for Bose
condensation. We confirm as in previous reports  that under high
energy non resonant excitation, the polariton laser is not obtained
and that only conventional photon lasing occurs. Nevertheless we
show that because of refraction index changes when increasing the
carrier density, this lasing regime can occur at much lower energy
that the uncoupled cavity mode deduced from low density
measurements. Thus in large Rabi splitting samples, a small emission
blueshift or even the absence of any blueshift does not prove the
persistence of the strong coupling. To demonstrate it one needs to
monitor the polariton dispersion while being aware that a high
angular resolution means that the emission from a large spatial area
is collected. Great care should be taken in analyzing angular
resolved PL in particular in spatially inhomogeneous geometries.
Finally we want to underline that the emission pattern of the photon
laser is the very same than that of a polariton condensate
co-existing with a thermalized population of uncondensed polaritons.

In our opinion, an unambiguous proof for polariton condensation or
polariton lasing is the observation of a second threshold at higher
excitation density\cite{Dang98,ThèseJacek,Bajoni07} corresponding to
photon lasing. This way it could be unambiguously shown that
polariton condensation occurs at low excitation while at higher
carrier density, the strong coupling regime vanishes and the onset
of photon lasing is observed.

We thank Esther Wertz for careful reading of the manuscript. We
acknowledge the European Community for funding through the Marie
Curie project ``Clermont 2" contract number MRTN-CT-2003-503677.
This work was partly supported by C'nano Ile de France, by the
"R\'egion Ile de France", and by the "Conseil G\'en\'eral de
l'Essonne".

\end{document}